\documentclass[aps,prc,twocolumn,floatfix,12pts,superscriptaddress]{revtex4}
\usepackage{epsfig}
\usepackage{amssymb}
\usepackage{amsmath}
\usepackage{color}
\usepackage{graphicx}
\usepackage{epsfig}
\usepackage{amsmath}
\usepackage{color}
\usepackage{graphicx}
\usepackage{float}
\usepackage{hyperref}
\definecolor{blue}{rgb}{0.05, 0.05, 0.5}

\begin{document}
\title{Probing Initial State Clustering through Photon Anisotropic Flow in 7A  TeV $^{16}$O+$^{16}$O Collisions at the LHC}

\author{Sanchari Thakur}
\email{sanchari.thakur@vecc.gov.in}
\affiliation{Department of Physical Sciences, Bose Institute, Kolkata, India}

\affiliation{Variable Energy Cyclotron Centre, 1/AF, Bidhan Nagar, Kolkata 700064, India}

\author{Pingal Dasgupta}
\email{dasgupta.pingal@ttk.elte.hu}

\affiliation{E\"{o}tv\"{o}s Lor\'{a}nd University, Department of Atomic Physics, Budapest, Hungary}

\author{Rupa Chatterjee}
\email{rupa@vecc.gov.in}

\affiliation{Variable Energy Cyclotron Centre, HBNI, 1/AF, Bidhan Nagar, Kolkata 700064, India}
\affiliation{Homi Bhabha National Institute, Training School Complex, Anushaktinagar, Mumbai 400094, India}

\author{Sinjini Chandra}

\affiliation{Variable Energy Cyclotron Centre, HBNI, 1/AF, Bidhan Nagar, Kolkata 700064, India}
\affiliation{Homi Bhabha National Institute, Training School Complex, Anushaktinagar, Mumbai 400094, India}

\author{Sidharth K. Prasad}
\affiliation{Department of Physical Sciences, Bose Institute, Kolkata, India}

\begin{abstract}
The presence of $\alpha$ clustered structures in light nuclei can enhance the initial spatial anisotropies in relativistic nuclear collisions relative to those arising from nuclei with uniform density distributions. Thus, observables that are strongly sensitive to the initial geometry can be a more efficient probe of the clustered structures than observables dominated by final state dynamics.
We investigate the collisions of $\alpha$ clustered oxygen nuclei at $\sqrt{s_{NN}}=7$A TeV at the LHC using the GLISSANDO initial state model along with the MUSIC event-by-event hydrodynamical framework. The tetrahedral $\alpha$ clustered structure of $^{16}$O leads to significantly larger initial triangular eccentricity $\epsilon_3$ than collisions with uniform density distributions especially in the most central events. 
The spatial eccentricity $\epsilon_2$ is found to be relatively less sensitive to the initial state clustered structure. The production of thermal photons is estimated to be only marginally influenced by clustering for  both central as well as peripheral collisions. 
In contrast, the photon triangular flow coefficient $v_3(p_T)$ is strongly affected by initial state clustering resulting in substantially larger values in both central and peripheral collisions. An experimental determination of photon anisotropic flow together with the ratios of flow coefficients in $^{16}$O+$^{16}$O collisions therefore expected to provide valuable insight into the possible clustered structure in light nuclei and also to constrain parameters in theoretical modeling.
\end{abstract}


\maketitle

\section{Introduction}

Alpha clustering is a well known phenomenon studied in nuclear structure research where the formation of $\alpha$ substructures is known to enhance stability in light nuclei~\cite{alpha1, alpha2, alpha3,alpha4}. Recent  studies suggest that the presence of $\alpha$ clustered structures in light nuclei, such as in $^{12}$C and $^{16}$O can be effectively investigated at relativistic energies by measuring the anisotropic flow parameters~\cite{cluster, bozek1, cluster1, cluster2, cluster3, cluster4, cluster5, song, sadhna_oo, china_oo, raghu, raghu1, raghu2, raghu3, glauber_oo}.

The clustered configurations of light nuclei lead to  pronounced spatial anisotropies in the overlap region when they collide with a  heavy nuclei at relativistic energies. In addition, collisions involving two clustered nuclei are also expected to significantly influence the initial energy density distribution on the transverse plane as well as the spatial anisotropy of the overlapping zone. Since the emission of electromagnetic radiation is expected to be  sensitive to the initial state,  the production and  anisotropic flow of direct photons can be a promising probe for studying the $\alpha$ clustered structures~\cite{Chatterjee:2005de, Srivastava:2008es, Chatterjee:2008tp, Chatterjee:2012dn, Chatterjee:2013naa, lhc_tau0, Liu:2012ax, Monnai:2014kqa, McLerran:2014hza, Vujanovic:2014xva, Gale:2014dfa, Chatterjee:2021gwa, phot_rev, ratio, uu, cluster_cau, cluster_ebye}. Additionally, the initial geometry driven effects are expected to be more prominent in the most central collisions compared to the peripheral ones~\cite{cluster_cau}.

It has been demonstrated that collisions of triangular $\alpha$ clustered $^{12}$C nuclei with Au at the top RHIC energy generate distinct initial geometries in the transverse plane depending on the orientations of the incoming clustered nuclei in central collisions~\cite{cluster_cau}. The initial anisotropic energy density distribution in these  collisions lead to a significantly enhanced anisotropic flow ($v_n(p_T)$) of thermal photons which shows strong dependence on the orientation angle.  In contrast, the $\alpha$ clustered structure was found to have only a marginal effect on the thermal photon spectra obtained at different orientations of C+Au collisions.

A more realistic calculation shows that the orientation averaged anisotropic flow of thermal photons in clustered $^{12}$C+Au collisions at 200A GeV exhibits significant qualitative differences between  the $v_2$ and $v_3$ parameters. The initial triangular eccentricity $\epsilon_3$ in the case of clustered $^{12}$C is estimated to be substantially larger compared to that in unclustered $^{12}$C+Au collisions~\cite{cluster_ebye}. This enhancement translates into a significantly larger triangular flow parameter $v_3(p_T)$ of photons for the clustered case. In contrast, the elliptic flow parameter is observed to be much smaller than the triangular flow in the clustered collisions.

The recent experimental run from $^{16}$O+$^{16}$O collisions at the LHC is expected to provide valuable insights into the formation of a hot and dense medium in small systems as well as into the role of clustered structures in light nuclei~\cite{alice_oo, opportunity, cms, atlas, star}. These results will also help determine whether investigating clustered structures requires collisions of light nuclei with heavy ions or if collisions between two light clustered nuclei are sufficient to reveal clear signatures of the clustered configuration in the initial state.

In the present work, we study the evolution of $\alpha$ clustered $^{16}$O+$^{16}$O collisions at 7A TeV at LHC using MUSIC hydrodynamical model framework~\cite{m1}. The GLISSANDO initial conditions is used  to construct the initial state for clustered and unclustered  collisions of oxygen nuclei~\cite{G1}.  We use the state of the art photon rates  to estimate the thermal photon production and anisotropic flow parameters for two different centrality bins at the LHC energy.

\section{Formalism}
We use GLISSANDO (Glauber Initial State Simulation AND mOre), a Monte Carlo-based framework to simulate the initial state in $\alpha$ clustered O+O collisions at $\sqrt{s_{NN}} = 7$A TeV at the LHC~\cite{G1, G2, G3}. The framework provides the spatial distributions of participant nucleons in different centrality bins for both clustered and unclustered configurations of $^{16}$O.


\begin{figure}
\centerline{\includegraphics*[width=8.4 cm]{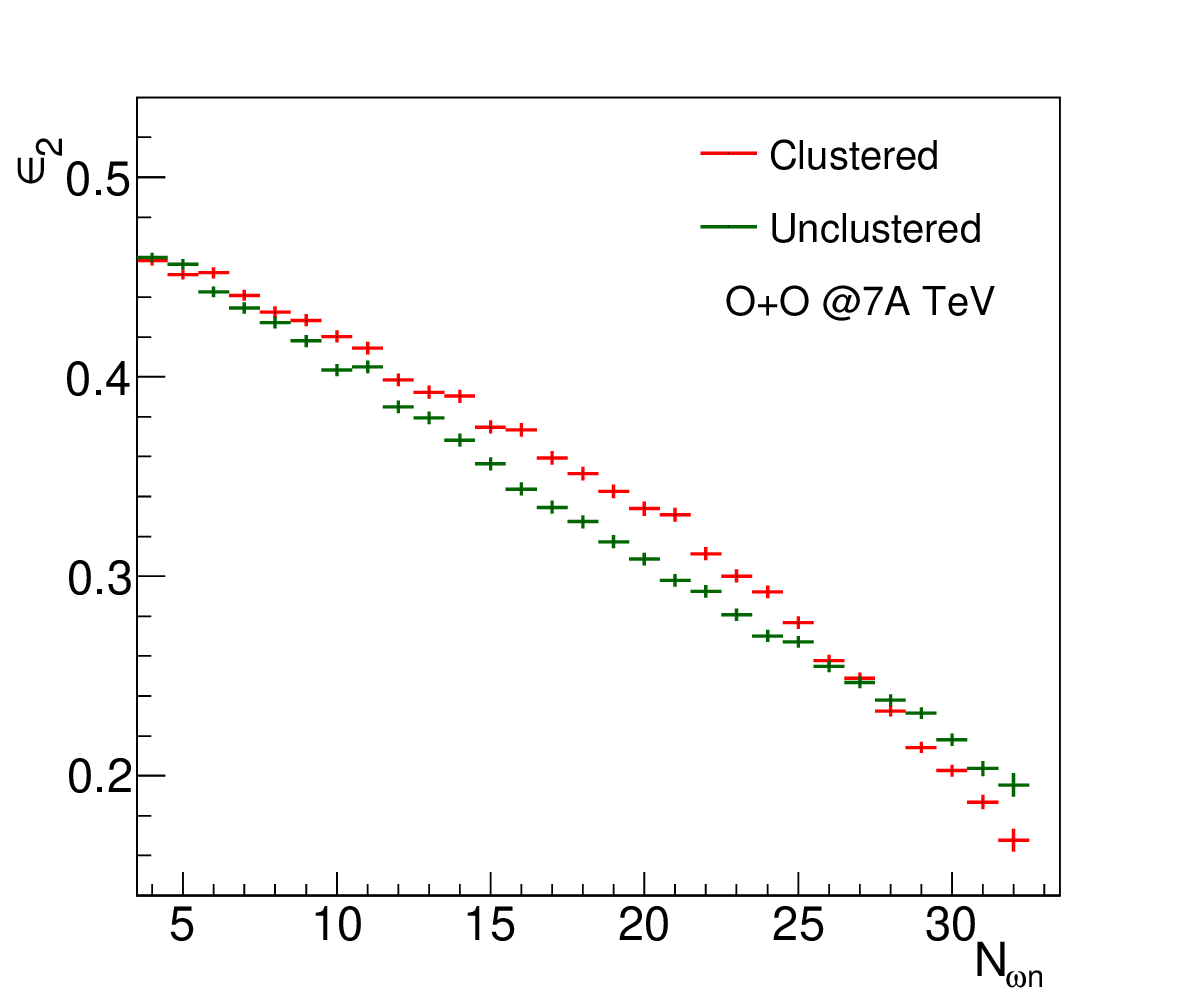}}
\caption{(Color online) Initial spatial eccentricity $\epsilon_2$ for clustered and unclustered O+O collisions at 7A TeV at LHC as a function of number of wounded nucleons $N_{\rm WN}$. }
\label{eps2}
\end{figure}

\begin{figure}
\centerline{\includegraphics*[width=8.4 cm]{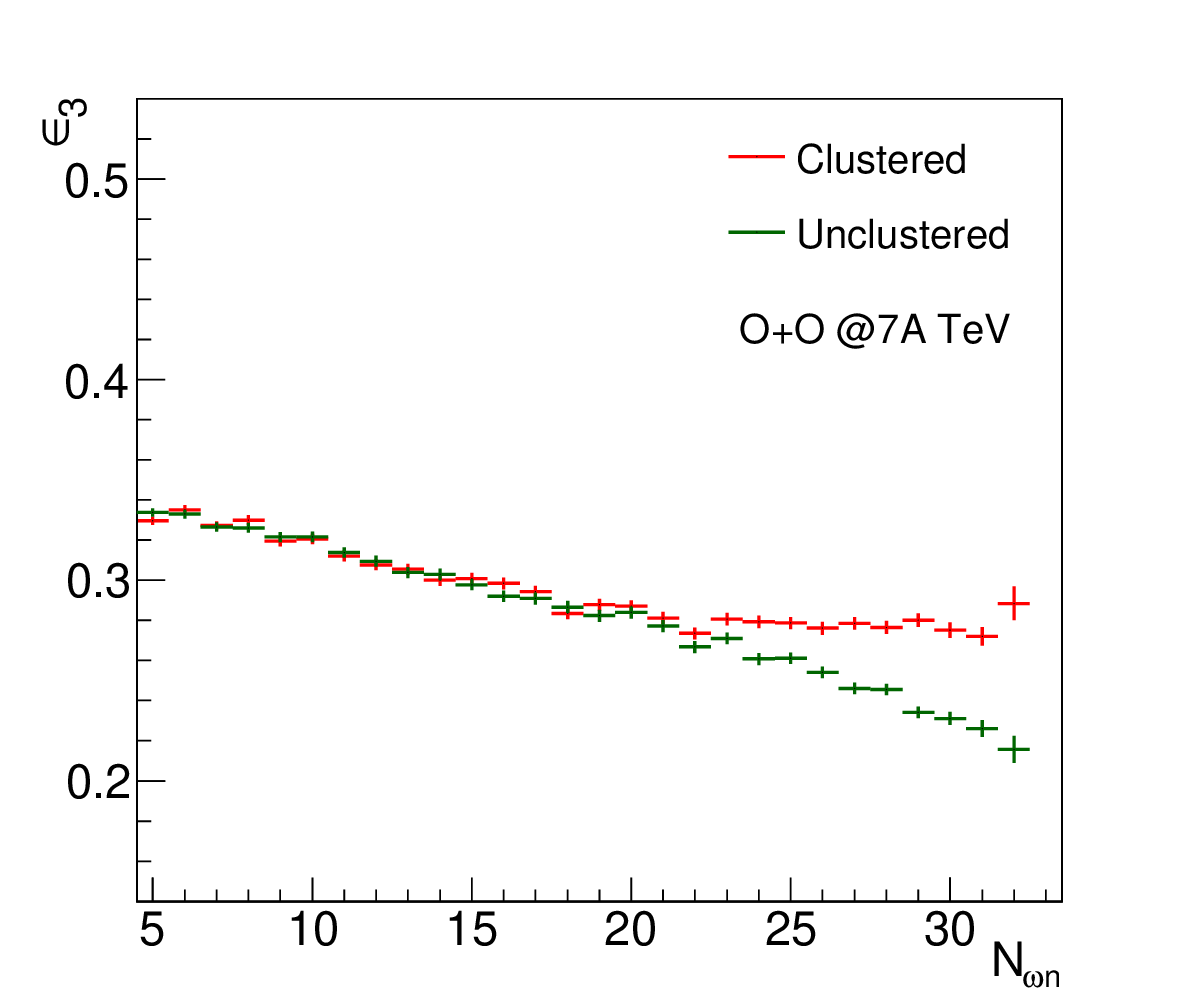}}
\caption{(Color online) Initial triangular eccentricity $\epsilon_3$ for clustered and unclustered O+O collisions at 7A TeV at LHC as a function of number of wounded nucleons $N_{\rm WN}$.}
\label{eps3}
\end{figure}
For the clustered configuration, the $^{16}$O  nucleus is modeled as a tetrahedral arrangement of four $\alpha$ clusters, with each cluster positioned at a vertex of the tetrahedron. The side length $l$ of the tetrahedron is taken as $3.2$ fm and the radius of each cluster $r_\alpha$ is  $1.1$ fm~\cite{o_structure, o_structure_1}.  The distribution of nucleons in each cluster is expressed through a Gaussian function as follows:
\begin{eqnarray}
f_i(\vec{r})=A \exp \left (- \frac{3}{2} \, (\vec{r}-\vec{c_i})^2/r_\alpha^2
\right ) \ .
\label{alpha_dist}
\end{eqnarray}
 Here, $\vec{c_i}$ denotes the position of the center of $i$th cluster in oxygen. The distance between the centers of any nucleon-nucleon pair is not to be less than 0.9 fm due to the short-range repulsion~\cite{bozek1}. The radial distribution of nucleon centers is used as a constraint when constructing the clustered configurations. For the unclustered case, an isotropic nuclear distribution whose root mean square (rms) radius is set equal to that of the clustered oxygen ($\sim$2.4 fm).

A two-component Monte Carlo Glauber model is used to determine the initial entropy density on the transverse plane~\cite{cluster_ebye}:
\begin{eqnarray}
s(x,y)=K {\sum_{i,j=1}^{N_{\rm part},N_{\rm coll} }}[ (1-\alpha)n_{\text{part}}(x_i,y_i)\ F_{i}(x,y) +\nonumber\\
\alpha \ n_{\text{coll}}(x_j,y_j)\ F_{j}(x,y) ]\ .
\label{imple_eqn}
\end{eqnarray}
In the above equation, $n_{\rm part}(x,y)$ and $n_{\rm coll}(x,y)$ denote the number of participants and the number of collisions at the position $(x,y)$ respectively. A participant is given a weight of (1 $-$ $\alpha$) and a binary collision is given a weight of $\alpha$, where $\alpha$ is taken as 0.12. The function $F_{i}(x,y)$ ($F_{j}(x,y)$) is a normalized Gaussian distribution function for the  $i^{th}$ ($j^{th}$) participant (collision) source where,
\begin{equation}
\label{normal_dist_eq}
F_{i,j}(x,y)=\frac{1}{2\pi \sigma ^2}\, e^{ -\frac{(x-x_{i,j})^2+(y-y_{i,j})^2}{2\sigma ^2}}.
\end{equation}
The smearing radius around each participant source ($\sigma$) is considered as $0.4$ fm~\cite{enhancement, cluster_ebye} and the overall normalization constant K in Eqn. (2) is adjusted to produce charged particle multiplicity ($dn_{\rm ch}/d\eta$) for most central O+O collisions $\sim$180 ~\cite{raghu}.

The initial state for each random event is then subsequently evolved with a longitudinally boost invariant (2+1) ideal relativistic hydrodynamic framework to obtain the space-time evolutions at the mid rapidity.  The MUSIC hydrodynamical framework has been used for this study, which has been used extensively to reproduce the charged particle spectra and anisotropic flow in relativistic nuclear collisions~\cite{m2, m3, m4}.  The initial formation time $\tau_0$  of the plasma is taken as 0.14 fm/$c$ and a lattice based EOS from Ref~\cite{eos} is used. The $\tau_0$ is taken as the same for Pb+Pb collisions at 2.76A TeV~\cite{lhc_tau0}. 
A constant freeze-out temperature of 137 MeV is considered for both the clustered and unclustered collisions and impact parameter range 0--1.37 and 2.6--3.0 fm respectively for  0--5\% and 30--40\% collision  centrality bins.

The production of thermal photons is estimated by integrating the  emission rates ($R=E\frac{dN}{d^3pd^4x}$) from QGP and hot hadronic matter over the space-time history:
        \begin{eqnarray}
        {E \frac{dN}{d^3p} \ =  \int \it{R}\big(E^*(x),T(x)\big)d^4x} \ .
        \label{dn_phot}
\end{eqnarray}
$T(x)$ in the above equation is the local temperature. $E^*(x)=p^\mu u_\mu(x)$, where $p^\mu$  represents the 4-momentum of photon and $u_\mu$ is the local 4-velocity of the flow field. 
We have used the NLO QGP rates from Refs.~\cite{amy, jacopo} and parametrized rates of photon production from Ref.~\cite{trg} to estimate the production and anisotropic flow of thermal photons at 7A TeV at the LHC. 

The JETPHOX framework has been employed to estimate NLO pQCD photon production at $\sqrt{s_{\rm NN}} = 7$ TeV~\cite{prompt}. The calculations use the CT14  parton distribution function~\cite{lhapdf} together with EPS09~\cite{eps09} parametrization of nuclear shadowing and the BFGII fragmentation function~\cite{frag}. The renormalization, factorization, and fragmentation scales are all set to $p_T$. An isolation cone of radius $R = 0.4$ and a transverse momentum cutoff of $p_T^{\rm cut} \approx 0.1p_T$ are applied to evaluate the total prompt photon production~\cite{Alice7TeV}. Results for the 0–-5\% and 30–-40\% centrality classes are obtained by scaling with the corresponding numbers of binary collisions taken to be approximately 53 and 24 respectively.
\begin{figure}
\centerline{\includegraphics*[width=8.4 cm]{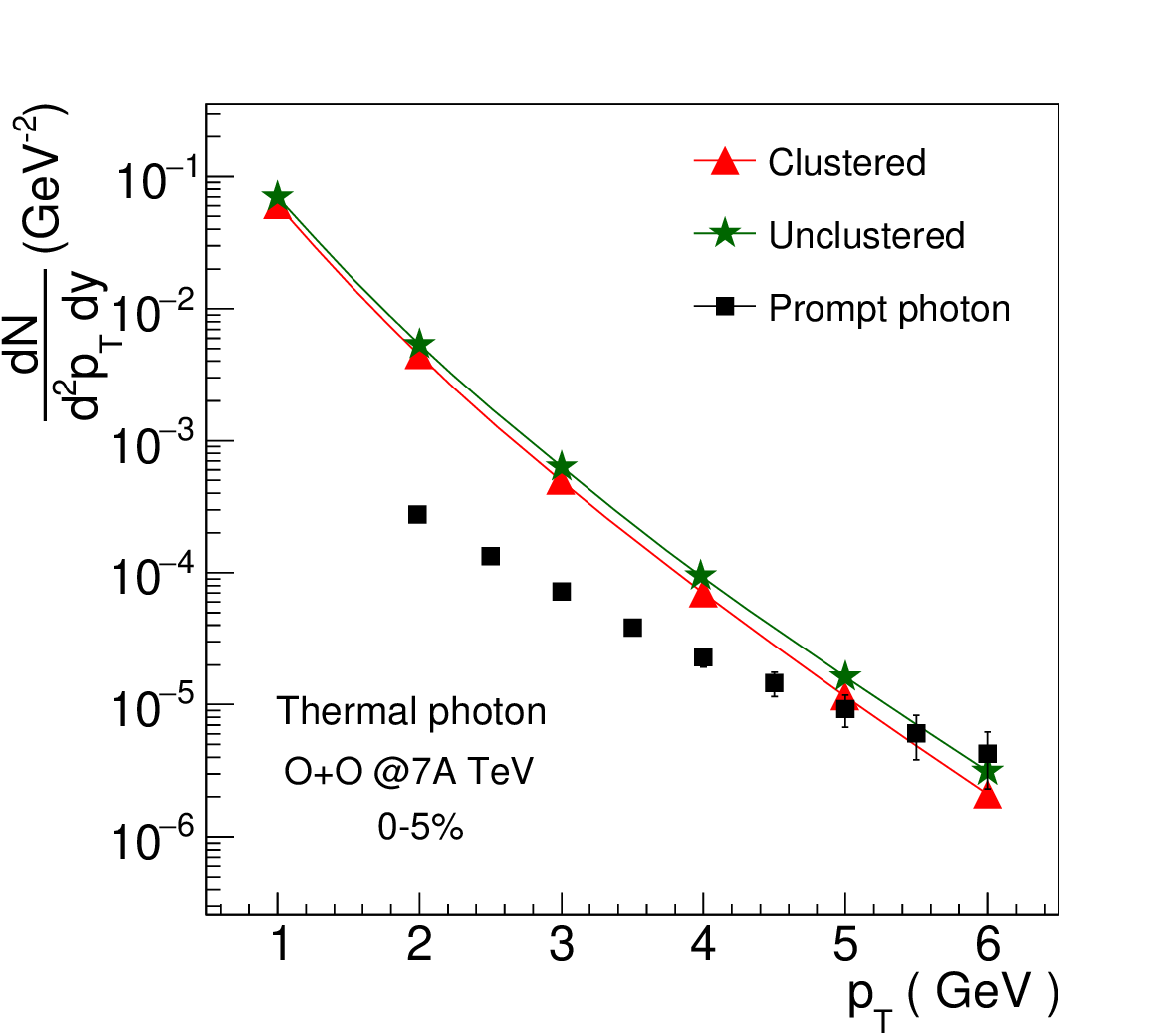}}
\centerline{\includegraphics*[width=8.4 cm]{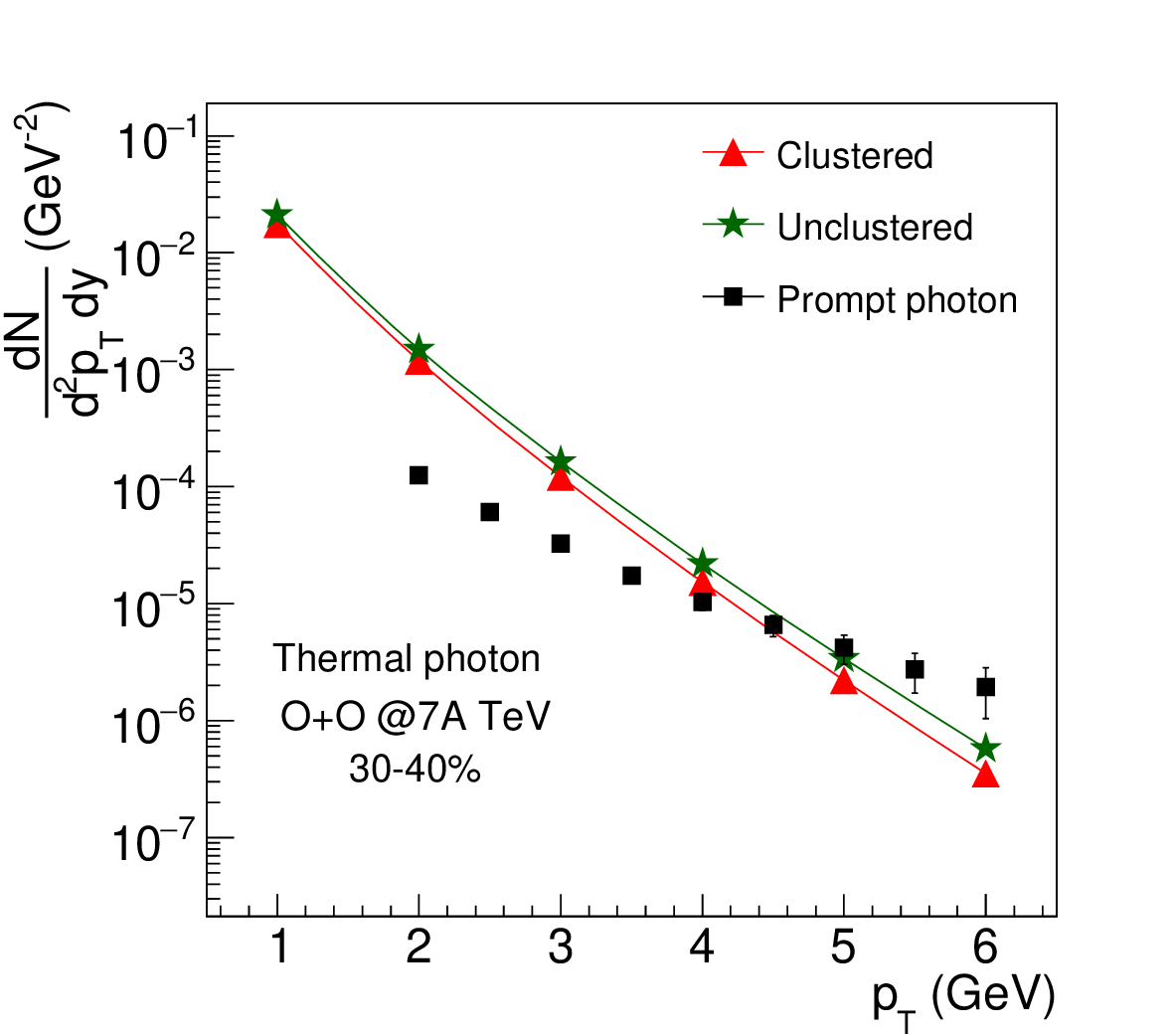}}
\caption{(Color online) Thermal photon spectra from 7A TeV O+O collisions at LHC for 0--5\% [upper panel] and 30--40\% [lower panel] centrality bins. The prompt photon spectra for the same centrality bins are shown for a comparison. }
\label{spec}
\end{figure}

The differential anisotropic flow coefficients $v_n(p_T)$ ($n=2$ and $3$) for each event is obtained as:
\begin{eqnarray}
v_n (p_T) \ = \ \frac{\int_0^{2\pi} \,
        d\phi\,{\rm cos} [\,n (\phi-\psi_{n})]\,\frac{dN}{p_T dp_T dy d\phi}}{\int_0^{2\pi}\,d\phi\,\frac{dN}{p_Tdp_Tdy d\phi}} \ .
\label{v2flow}
\end{eqnarray}
Where, $\psi_n$ is the participant plane angle  and   $\phi$ is the azimuthal angle of particle momentum. The participant plane angle is estimated using the relation,
\begin{eqnarray}
\psi_{n} = \frac{1}{n} \arctan
\frac{\int \mathrm{d}x \mathrm{d}y \; r^2 \sin \left( n\Phi \right) \epsilon\left( x,y,\tau _{0}\right) }
{ \int \mathrm{d}x \mathrm{d}y \; r^2 \cos \left( n\Phi \right) \epsilon\left( x,y,\tau _{0}\right)}  + \pi/n \, ,
\label{sai_nm}
\end{eqnarray}
It is to be noted that the photon anisotropic flow results from event plane methods are found to be similar to those obtained from the participant plane method~\cite{cluster_ebye}.
Additionally, the inclusion of viscosity is essential for a more quantitative comparison with experimental data. However, the qualitative features of the results due to the presence of $\alpha$ clustered structures are expected to remain unchanged under viscous hydrodynamic expansion.

\section{Results and discussion}
The initial spatial eccentricities $\epsilon_2$ and $\epsilon_3$ for $\alpha$ clustered and unclustered O+O collisions at 7A TeV at the LHC as a function of wounded nucleons ($N_{\rm wn}$) are shown in Figs.~\ref{eps2} and~\ref{eps3} respectively. The spatial eccentricities are obtained by averaging over $10^6$ random events for each configuration. The $\epsilon_2$ is found to be small but nonzero even for the most central collisions for both cases arising from initial state fluctuations. In addition,  the $\epsilon_2$ in the most central collisions is smaller for the clustered case than the unclustered one, reflecting the reduced role of fluctuations in the presence of a correlated clustered structure. Whereas, in mid-central collisions, the $\epsilon_2$ is slightly larger for the clustered collisions than for the unclustered collisions.

\begin{figure}
\centerline{\includegraphics*[width=8.4 cm]{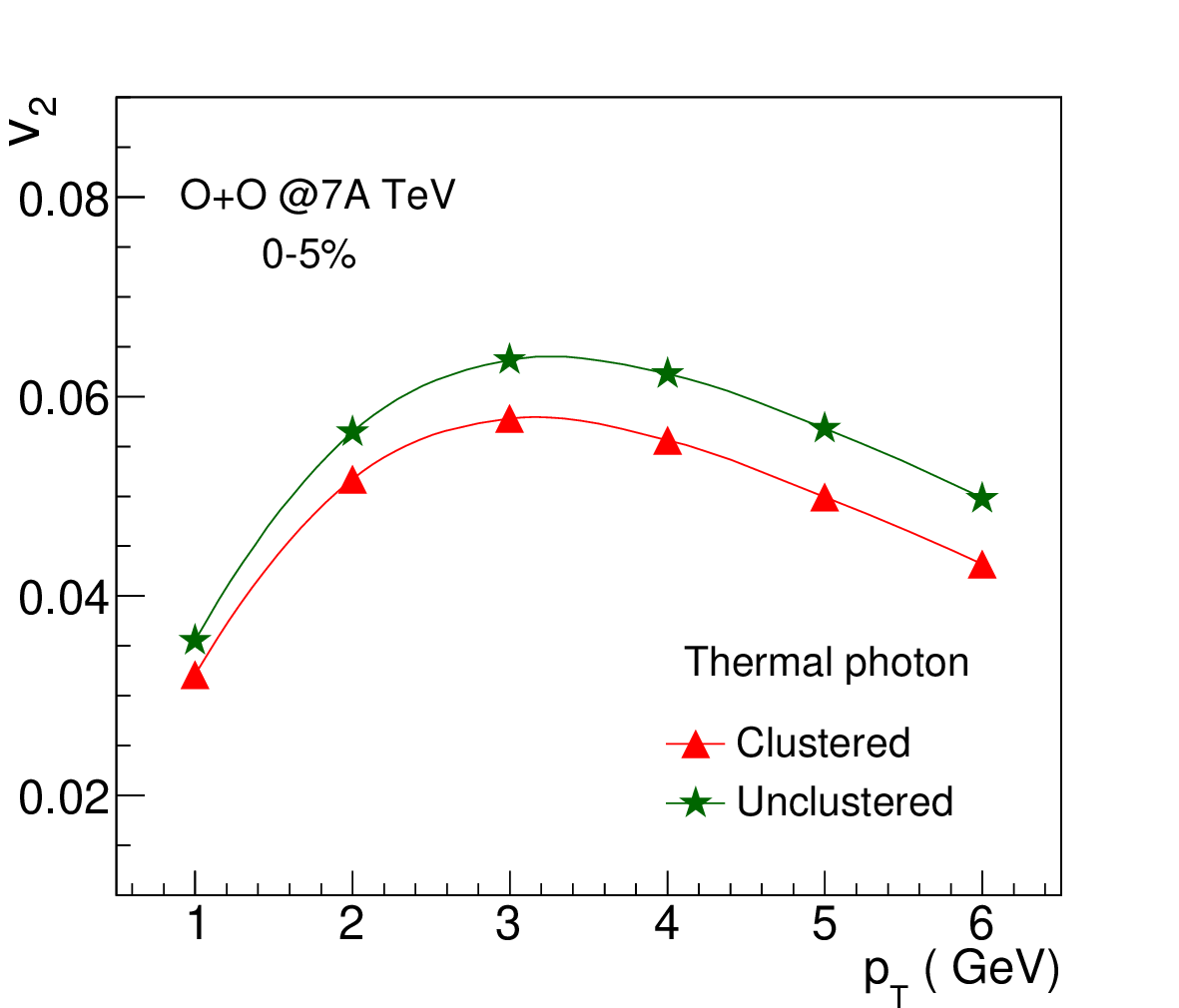}}
\caption{(Color online) Elliptic flow of thermal photons from clustered and unclustered O+O collisions at 7A TeV at LHC for  0--5\% centrality bin. }
\label{figv2_5}
\end{figure}

\begin{figure}
\centerline{\includegraphics*[width=8.4 cm]{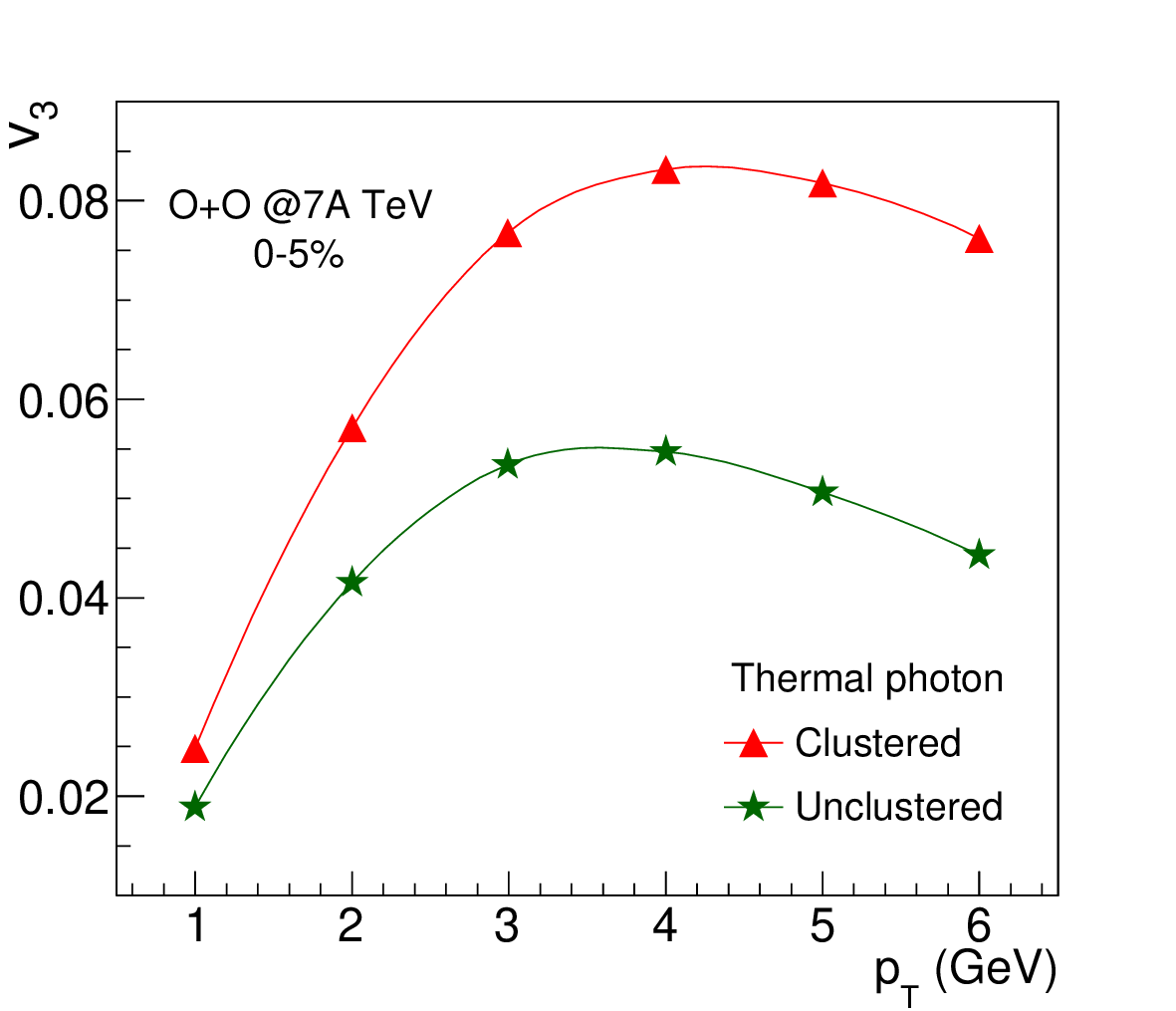}}
\caption{(Color online) Triangular flow of thermal photons from clustered and unclustered O+O collisions at 7A TeV at LHC for  0--5\% centrality bin. }
\label{figv3_5}
\end{figure}

The value of $\epsilon_3$ is found to be significantly larger for the clustered configuration compared to the unclustered one for central collisions [see Fig.~\ref{eps3}]. In contrast, for mid-central and peripheral collisions, the initial triangular eccentricity shows no noticeable dependence on the clustered structure.

The $\epsilon_3$  exhibits weaker dependence to collision centrality compared to $\epsilon_2$  except for very peripheral events.

\begin{figure}
\centerline{\includegraphics*[width=8.4 cm]{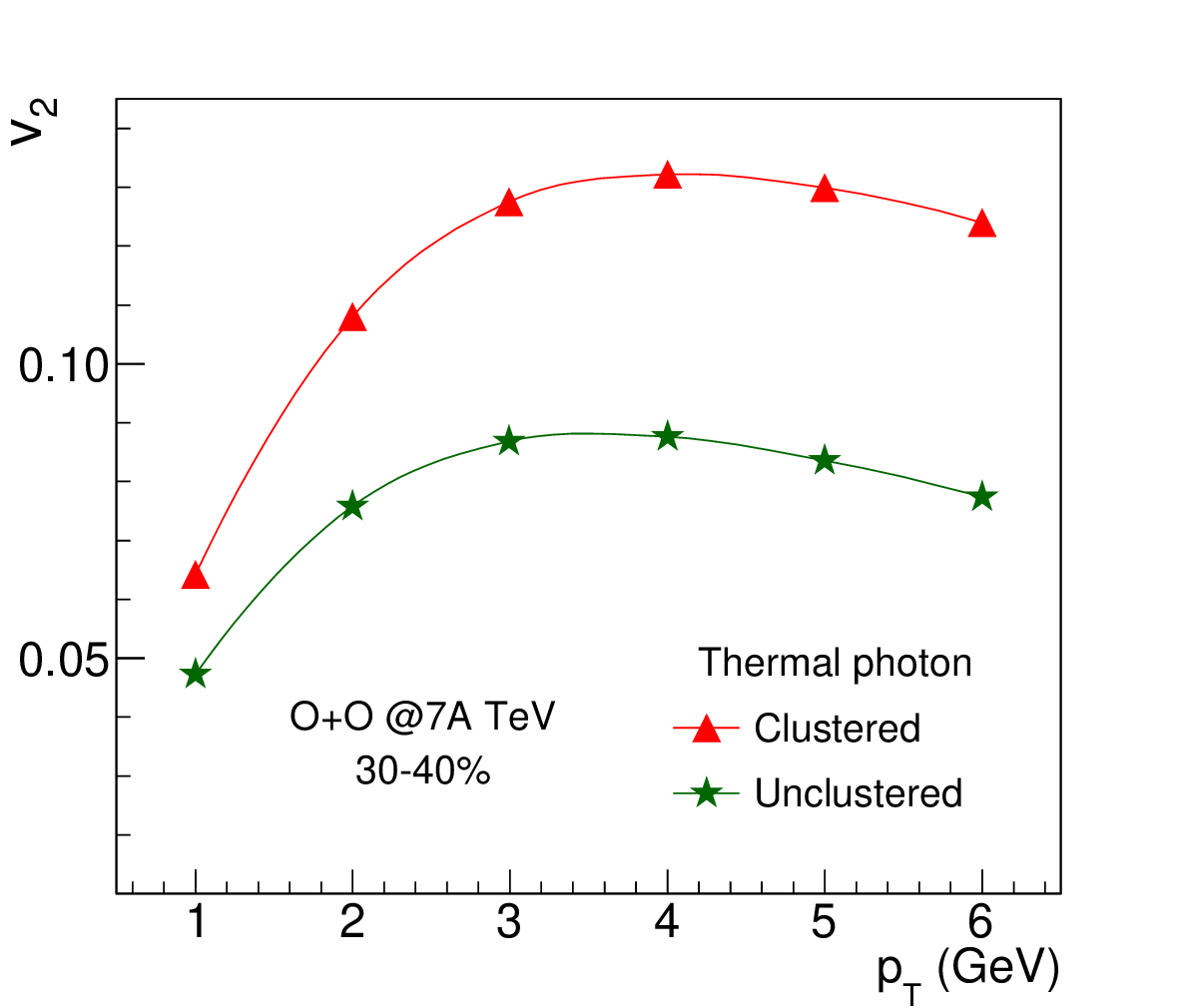}}
\caption{(Color online) Elliptic flow of thermal photons from clustered and unclustered O+O collisions at 7A TeV at LHC for  30--40\% centrality bin.}
\label{figv2_30}
\end{figure}

\begin{figure}
\centerline{\includegraphics*[width=8.4 cm]{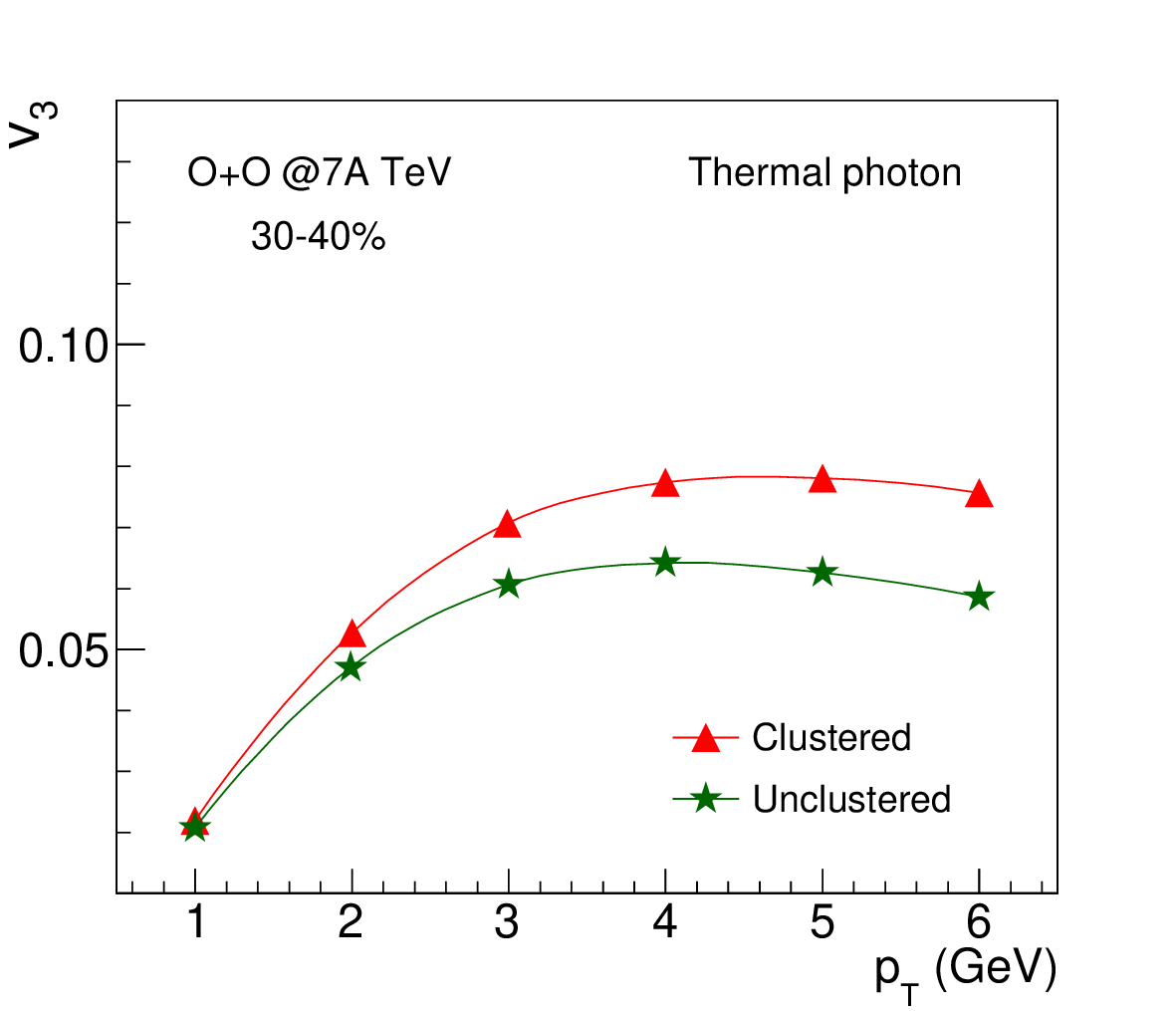}}
\caption{(Color online) Triangular flow of thermal photons from clustered and unclustered O+O collisions at 7A TeV at LHC for  30--40\% centrality bin. }
\label{figv3_30}
\end{figure}

\subsection{Photon spectra}
The thermal photon spectra from clustered and unclustered O+O collisions at 7A TeV are shown in Figs.~\ref{spec}.  The thermal spectra are estimated for two centrality bins, 0–-5\% and 30–-40\%. The prompt photon spectra for both centrality bins are shown for a comparison.

The presence of clustered structure in oxygen nuclei is found to have only a marginal effect on photon production in the low $p_T$ (below 3 GeV) region for both central and peripheral collisions. The effect of initial state fluctuations is expected to be more pronounced for larger $p_T$ region, which might result in a small difference in the spectra from clustered and unclustered cases for $p_T \ >$ 3 GeV.

The fluctuations may be more significant in the unclustered case due to the random distribution of nucleons in the small volume, which may lead to an enhanced photon production relative to clustered collisions.

\subsection{Anisotropic flow of thermal photons}
The anisotropic flow of thermal photons from 0--5\% central O+O collisions at 7A TeV at the LHC is shown in Figs.~\ref{figv2_5} and~\ref{figv3_5} for both clustered and unclustered collisions. The anisotropic flow results are obtained by taking an average over 400 random events. The elliptic flow parameter as a function of $p_T$ shows similar nature to the $v_2(p_T)$ of photons from heavy ion collisions at RHIC and LHC energies~\cite{Chatterjee:2021gwa}.

For the 0–-5\% centrality bin, the elliptic flow is found to be significantly larger in the unclustered case due to initial state fluctuations. In contrast, the presence of a clustered configuration suppresses these fluctuations due to correlations among the clustered sites. Consequently, both the initial eccentricity $\epsilon_2$ and the resulting elliptic flow $v_2(p_T)$ are found to be smaller in the clustered case compared to the unclustered scenario.

The elliptic and triangular flow exhibit distinct qualitative behavior for the clustered and unclustered configurations for 0--5\% centrality bin. The $p_T$ dependent triangular flow parameter is significantly larger than the elliptic flow coefficient in the clustered configuration. However,  for the unclustered (or uniform) case the elliptic flow is considerably larger than the triangular flow $v_3(p_T)$. These anisotropic flow patterns can be understood from the eccentricity plots presented earlier, the larger $\epsilon_3$ in the clustered case and the relatively larger $\epsilon_2$ in the unclustered case primarily decide the observed anisotropic flow results.

For the peripheral collisions of 30--40\% centrality bin, the elliptic as well as triangular flow are both found to be larger for the clustered configuration than unclustered collisions [Figs.~\ref{figv2_30} and~\ref{figv3_30}].

Additionally, the $v_4(p_T)$ coefficient of thermal photons is found to be negligible even though the initial $\epsilon_4$ is significantly large for both the collision centralities [not shown here]. Thus, the tetrahedral $\alpha$ clustered structure in the oxygen nucleus is found to affect the triangular flow parameter maximum. 

\subsection{Ratio of $v_2$ and $v_3$}
\begin{figure}
\centerline{\includegraphics*[width=8.4 cm]{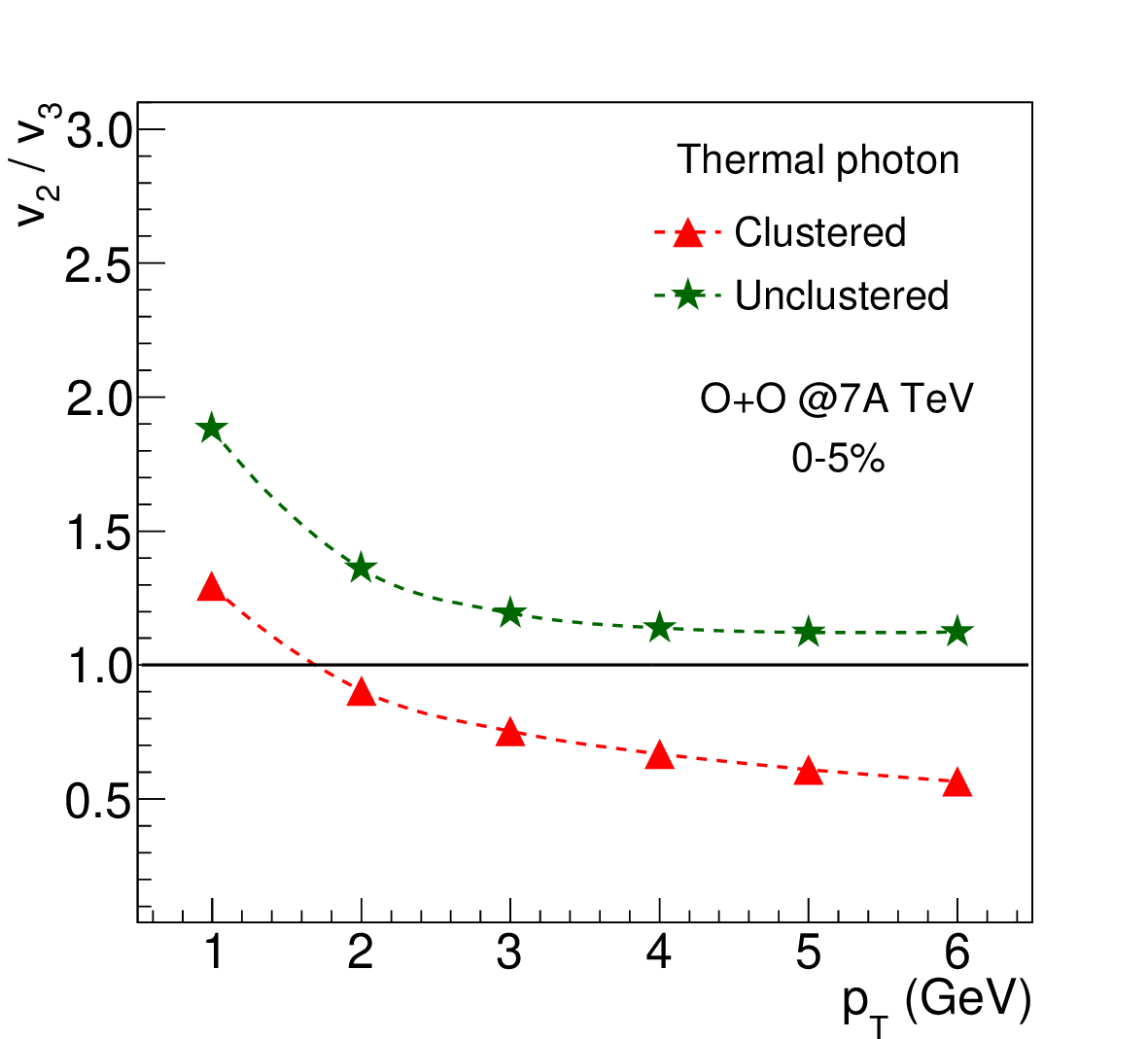}}

\caption{(Color online) Ratio of elliptic and triangular flow of thermal photons in clustered and unclustered O+O collisions at 7A TeV at LHC for  0--5\% centrality bin. }
\label{ratio_5}
\end{figure}

\begin{figure}
\centerline{\includegraphics*[width=8.4 cm]{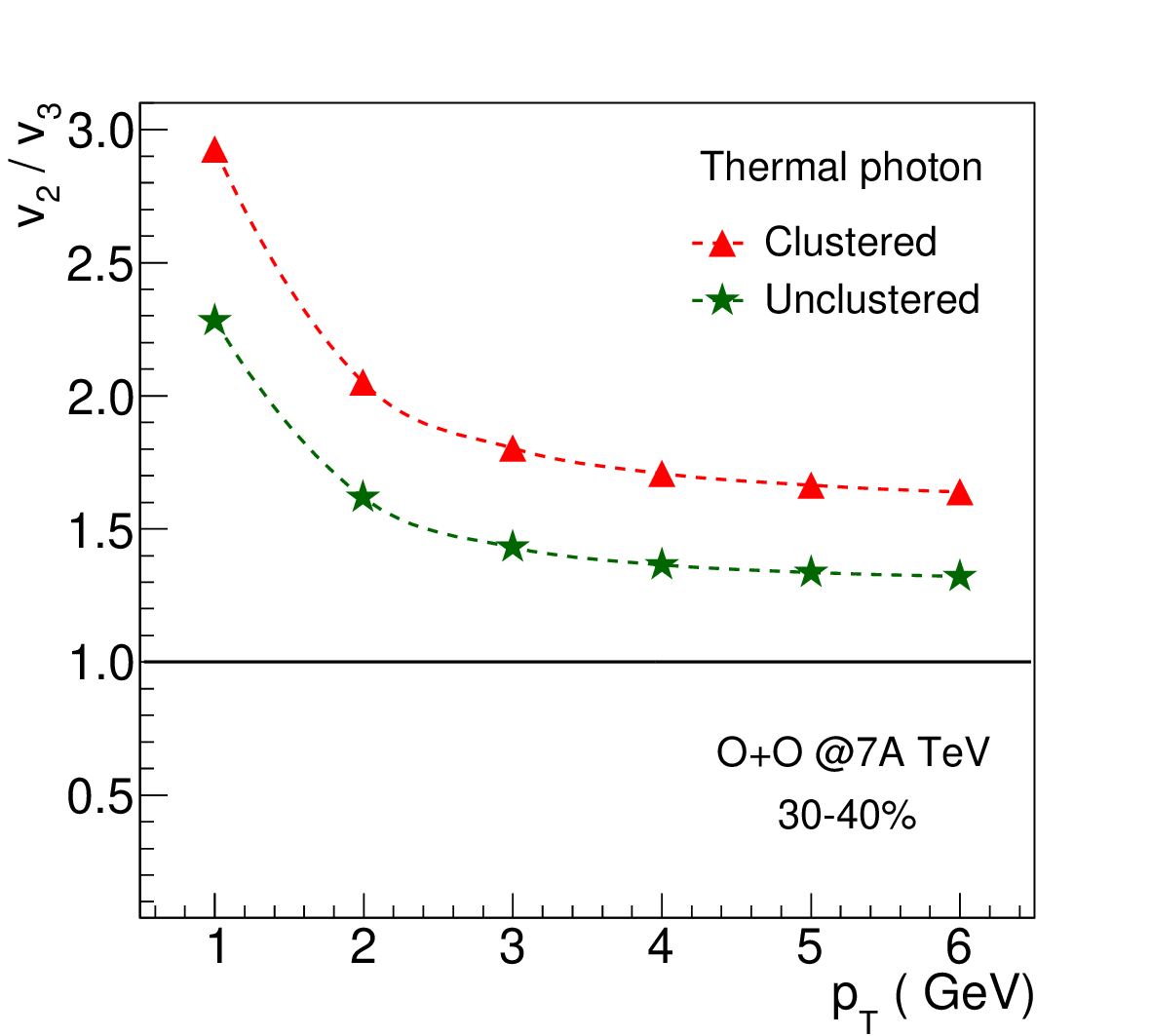}}

\caption{(Color online) Ratio of elliptic and triangular flow of thermal photons in clustered and unclustered O+O collisions at 7A TeV at LHC for  30--40\% centrality bin. }
\label{ratio_30}
\end{figure}

The ratio of the elliptic and triangular flow parameters ($v_2/v_3$) as a function of $p_T$ is shown in Fig.~\ref{ratio_5} and~\ref{ratio_30}. As expected from the individual $v_n$ results, the ratio from clustered collisions is smaller than 1 for $p_T \ >$ 1 GeV. Whereas, the same ratio is larger than 1 for the unclustered case in the entire $p_T$ region. The results from peripheral collisions show that the ratio is always greater than 1 as $v_2(p_T)$ is always larger than $v_3(p_T)$ for both configurations. 

\section{summary and conclusions}
The GLISSANDO initial conditions combined with the MUSIC hydrodynamical model have been used to study the initial state, photon production, and anisotropic flow parameters of thermal photons in 7A TeV O+O collisions at the LHC. 
The O+O collisions, being a small system, exhibit sizable initial state eccentricities from event-by-event fluctuations even in the most central events, leading to large $\epsilon_n$ ($n=2,3$) for both clustered and unclustered cases. These fluctuations are expected to be more pronounced in the unclustered configuration.

However, the presence of a tetrahedral $\alpha$ clustered structure in oxygen nuclei is found to significantly affect the initial triangular eccentricity in the most central collisions compared to the unclustered O+O collisions.

The photon spectra from clustered and unclustered configurations are found to be close to each other for both central and peripheral collisions. Only at $p_T \ >$ 3 GeV, the photon production from unclustered O+O collisions is found to be slightly larger than the clustered one.

The triangular flow of thermal photons in clustered O+O collisions is found to be significantly larger than $v_3(p_T)$ in the unclustered configuration for the most central collisions. In contrast, the elliptic flow in the unclustered case is slightly higher than that in clustered collisions. For peripheral collisions however, both anisotropic flow coefficients are larger in the $\alpha$ clustered configuration compared to the unclustered case. The elliptic flow shows a strong centrality dependence increasing substantially from central to peripheral collisions whereas, the triangular flow exhibits little sensitivity to collision centrality.

The ratio of photon $v_2$ and $v_3$ as a function of $p_T$ for different centrality bins is expected to exhibit qualitatively different behavior due to the presence of the clustered structure which could be verified experimentally.

\begin{acknowledgments}
We acknowledge the use of the GRID and Kanaad computing facilities at VECC for this study. We are also grateful to Mr. Abhishek Seal for his assistance and support with the GRID computing facility.
\end{acknowledgments}


\begin{thebibliography}{50}


\bibitem{alpha1} G. Gamow, Constitution of atomic nuclei and radioactivity (Oxford, 1931). 

\bibitem{alpha2} F. Hoyle, 
 \ Astrophys. \ J. \ Suppl. {\bf 1}, 121 (1954). 

\bibitem{alpha3}R. Bijker and F. Iachello,
\ Phys. \ Rev. \ Lett. {\bf 112}, 152501 (2014).

\bibitem{alpha4} D. DellAquila {\it et al.},
\ Phys. \ Rev. \ Lett. {\bf 119}, 132501 (2017). 

\bibitem{cluster} W. Broniowski and E. Ruiz Arriola, 
 \ Phys. \ Rev. \ Lett. {\bf 112}, 112501 (2014).

\bibitem{bozek1} P. Bozek, W. Broniowski, E. R. Arriola and M. Rybczynski, \ Phys. \ Rev. \ C {\bf 90}, 064902 (2014).

\bibitem{cluster1} S. Zhang, Y. G. Ma, J. H. Chen, W. B. He, and C. Zhong, 
\ Phys. \ Rev. \ C {\bf 95}, 064904 (2017).


\bibitem{cluster2} M. Rybczynski and W. Broniowski, 
 \ Phys. \ Rev. \ C  {\bf 100}, 064912 (2019).

\bibitem{cluster3} Y.-Z. Wang, S. Zhang, and Y.-G. Ma, 
 \ Phys.  \ Lett. \ B {\bf 831}, 137198 (2022).

\bibitem{cluster4} Y.-A. Li, S. Zhang, and Y.-G. Ma, 
\ Phys. \ Rev. \ C {\bf 102}, 054907 (2020).

        
\bibitem{cluster5}
J.~He, W.~B.~He, Y.~G.~Ma and S.~Zhang,
Phys. Rev. C \textbf{104}, 044902 (2021).

\bibitem{song} Y. Wang, S. Zhao, B. Cao, H.-J. Xu, and H. Song,
 \ Phys. \ Rev. \ C  {\bf 109}, L051904 (2024).


\bibitem{sadhna_oo}  
D. Sharma, A. Singh, Md. S. Islam, B. Nandi, and S. Dash, arXiv:2503.20339.

\bibitem{china_oo}
 C. Ding, L.-G. Pang,  S. Zhang, and Y.-G. Ma, \ Chin. \ Phys. \ C {\bf 47}, 024105  (2023).

 \bibitem{raghu1}
D. Behera, S. Prasad, N. Mallick, and R. Sahoo, \ Phys. \ Rev.  \ D {\bf 108}, 054022  (2023).

\bibitem{raghu2} S. Prasad, N. Mallick, R. Sahoo, and G. G. Barnafoldi, \ Phys. \ Lett. \ B {\bf 860}, 139145 (2025).

\bibitem{raghu3}
D. Behera, S. Deb, C. R. Singh, and R. Sahoo1, \ Phys. \ Rev. \ C {\bf 109}, 014902 (2024).

\bibitem{raghu}
D. Behera, N. Mallick, S. Tripathy, S. Prasad, and A. N. Mishra, and R. Sahoo,  \ Eur. \ Phys. \ J. \ A {\bf 58}, 157 (2022).

\bibitem{glauber_oo}
C. Loizides, arXiv:2507.05853.


\bibitem{Chatterjee:2005de}
R.~Chatterjee, E.~S.~Frodermann, U.~W.~Heinz and D.~K.~Srivastava,
Phys. Rev. Lett. \textbf{96}, 202302 (2006).

\bibitem{Srivastava:2008es}
D.~K.~Srivastava,
J. Phys. G \textbf{35}, 104026 (2008).
              
\bibitem{Chatterjee:2008tp}
R.~Chatterjee and D.~K.~Srivastava,
Phys. Rev. C \textbf{79}, 021901 (2009); R.~Chatterjee and D.~K.~Srivastava,
Nucl. Phys. A \textbf{830}, 503C-506C (2009).
                        

\bibitem{Chatterjee:2012dn}
R.~Chatterjee, H.~Holopainen, T.~Renk and K.~J.~Eskola,
Phys. Rev. C \textbf{85}, 064910 (2012).

                 
\bibitem{Chatterjee:2013naa}
R.~Chatterjee, H.~Holopainen, I.~Helenius, T.~Renk and K.~J.~Eskola,
Phys. Rev. C \textbf{88}, 034901 (2013).

  


\bibitem{lhc_tau0}
R.~Chatterjee, H.~Holopainen, T.~Renk and K.~J.~Eskola,
Phys. Rev. C \textbf{85} (2012), 064910.
                                                      
\bibitem{Liu:2012ax}
F.~M.~Liu and S.~X.~Liu,
Phys. Rev. C \textbf{89}, 034906 (2014).

\bibitem{Monnai:2014kqa}
A.~Monnai,
Phys. Rev. C \textbf{90},  021901 (2014).
                                              
\bibitem{McLerran:2014hza}
L.~McLerran and B.~Schenke,
Nucl. Phys. A \textbf{929}, 71 (2014).
                                             

  

\bibitem{Vujanovic:2014xva}
G.~Vujanovic, J.~F.~Paquet, G.~S.~Denicol, M.~Luzum, B.~Schenke, S.~Jeon and C.~Gale,
Nucl. Phys. A \textbf{932}, 230  (2014).

                            
\bibitem{Gale:2014dfa}
C.~Gale, Y.~Hidaka, S.~Jeon, S.~Lin, J.~F.~Paquet, R.~D.~Pisarski, D.~Satow, V.~V.~Skokov and G.~Vujanovic,
Phys. Rev. Lett. \textbf{114}, 072301 (2015).


\bibitem{Chatterjee:2021gwa}
R.~Chatterjee,
Pramana \textbf{95}, 15 (2021).

\bibitem{phot_rev}
D.~Blau and D.~Peresunko,
Particles \textbf{6}, 173 (2023).

\bibitem{ratio}
R. Chatterjee and P. Dasgupta,
 \ Phys. \ Rev. \ C {\bf 104}, 064907 (2021).


\bibitem{uu}
P.~Dasgupta, R.~Chatterjee and D.~K.~Srivastava,
Phys. Rev. C \textbf{95}, 064907 (2017).


 

\bibitem{cluster_cau}
P.~Dasgupta, G.~L.~Ma, R.~Chatterjee, L.~Yan, S.~Zhang and Y.~G.~Ma,
Eur. Phys. J. A \textbf{57}   134(2021).


\bibitem{cluster_ebye} 
P. Dasgupta, R. Chatterjee, G.-L. Ma,
\ Phys. \ Rev. \ C {\bf 107}, 4 (2023).


\bibitem{alice_oo}
 ALICE Collaboration,
ALICE-PUBLIC-2021-004.


\bibitem{opportunity} Opportunities of O+O and p+O Collisions at the LHC, CERN Workshop, Geneva, 2017, http://cern.ch/OppOatLHC.

\bibitem{cms}
CMS Collaboration,
CMS-PAS-HIN-25-008.

\bibitem{atlas}
 ATLAS Collaboration,
ATLAS-CONF-2025-010.


\bibitem{star}
S.~Huang,
[arXiv:2312.12167 [nucl-ex]].


\bibitem{m1} B. Schenke, S. Jeon, and C. Gale, \ Phys. \ Rev. \ C {\bf 82} 014903 (2010).

\bibitem{G1} 
W.~Broniowski, M.~Rybczynski and P.~Bozek,
 \ Comput. \ Phys. \ Commun. \textbf{180}, 69 (2009).




\bibitem{G2}
M.~Rybczynski, G.~Stefanek, W.~Broniowski and P.~Bozek,
\ Comput. \ Phys. \ Commun. \textbf{185}, 1749 (2014).


\bibitem{G3}
P.~Bo{\.z}ek, W.~Broniowski, M.~Rybczynski and G.~Stefanek,
 \ Comput. \ Phys. \ Commun. \textbf{245}, 106850 (2019).



\bibitem{o_structure} R. B. Wiringa, R. Schiavilla, S. C. Pieper, and J. Carlson,
\ Phys. \ Rev. \ C {\bf 89}, 024305 (2014).

\bibitem{o_structure_1}  D. Lonardoni, A. Lovato, S. C. Pieper, and R. B. Wiringa,
\ Phys. \ Rev. \ C {\bf 96}, 024326 (2017).


 
\bibitem{enhancement} R. Chatterjee, H. Holopainen, T. Renk, and K. J. Eskola, 
 \ Phys. \  Rev. \  C \ {\bf 83}, 054908 (2011). 
 
\bibitem{m2} \ Phys. \ Rev. \ Lett. {\bf 106}, 042301 (2011).


\bibitem{m3} J.-F. Paquet, C. Shen, G. S. Denicol, M. Luzum, B. Schenke, S. Jeon, and C. Gale, 
 \ Phys. \ Rev. \ C {\bf 93}, 044906 (2016).

\bibitem{m4} C. Shen, J.-F. Paquet, G. S. Denicol, S. Jeon, and C. Gale, \ Phys.\ Rev. \ C {\bf 95}, 014906 (2017).


\bibitem{eos} P. Huovinen, P. Petreczky, \ Nucl. \  Phys. \ A {\bf 837}, 26  (2010).

\bibitem{amy} P. B. Arnold, G. D. Moore and L. G. Yaffe,
 \ JHEP  \ {\bf 12}, 009 (2001).
 
\bibitem{jacopo} J. Ghiglieri, J. Hong, A. Kurkela, E. Lu, G. D. Moore
and D. Teaney, 
 \ JHEP {\bf 05}, 010 (2013).

 \bibitem{trg} S. Turbide, R. Rapp and C. Gale, 
 \ Phys. \ Rev. C {\bf 69}, 014903 (2004).


 


\bibitem{prompt} P. Aurenche, J.-P. Guillet, E. Pilon, and M. Werlen, \ Phys. \ Rev. \ D {\bf 73}, 094007 (2006).

\bibitem{lhapdf} A. Buckley {\it et. al.}, \ Eur. \ Phys. \ J. C {\bf 75} 3, 132 (2015).

\bibitem{eps09} K. J. Eskola, H. Paukkunen, C. A. Salgado, JHEP {\bf 04}, 065 (2009).

\bibitem{frag} L. Bourhis, M. Fontannaz and J. P. Guillet, \ Eur. \ Phys. \ J. \ C {\bf 2}, 529 (1998).

\bibitem{Alice7TeV} S. Acharya {\it et. al.,} [ALICE Collaboration], \ Eur. \ Phys. \ J. \ C {\bf 79}, 896 (2019).















  











 


\end{thebibliography}
\end{document}